\documentstyle[referee,epsfig,graphicx]{mn2e}
\newif\ifAMStwofonts
\newcommand{\be}{\begin{equation}}
\newcommand{\ee}{\end{equation}}
\newcommand{\bea}{\begin{eqnarray}}
\newcommand{\eea}{\end{eqnarray}}

\newcommand{\da}{\frac{\delta}{\alpha}}

\newcommand{\ad}{\frac{\alpha}{\delta}}

\title{ Isolated and non-isolated dark matter halos and the NFW profile }

\author[ R.N. Henriksen]
{
R.N. Henriksen$^1$\thanks{henriksn@astro,queensu.ca}
\\
$^1$Queen's University, Kingston, Ontario, K7L 3N6,Canada \\
and SAp/DAPNIA, CEA Saclay, 91191 Gif-sur Yvette, CEDEX, France\\
}
\date{\today}
\pagerange{\pageref{firstpage}--\pageref{lastpage}}
\pubyear{2005}

\begin{document}

\maketitle

\label{firstpage}

\begin{abstract}
   
We compare the conclusions reached using the coarse-graining technique employed  by Henriksen (2004) for a one degree of freedom (per particle) collisionless system, to those presented  in a paper by Binney (2004) based on an exact one degree of freedom model. We find  agreement in detail but in addition we show that the isolated 1D system is self-similar and therefore unrelaxed. Fine graining of this system recovers much less prominent wave-like structure than in a spherically symmetric isotropic 3D system. The rate of central flattening is also reduced in the 1D system. We take this to be evidence that relaxation of collisionless systems proceeds ultimately by way of short wavelength Landau damping.
N-body systems, both real and simulated, can  be trapped in an incompletely relaxed state because of a break in the cascade of energy to small scales. This  may be due to the rapid dissipation of the small scale oscillations in an isolated system, to the existence of conserved quantities such as angular momentum, or to the failure in simulations to resolve sub-Jeans length scales. Such a partially relaxed state appears to be the NFW state, and is to be expected especially in young systems. The NFW core is shown to be isolated.  In non-isolated systems continuing coarse-grained relaxation should be towards a density core in solid body rotation.

     \end{abstract} 
\setlength{\baselineskip}{13pt}
\section{Introduction}
\label{sec:intro}

The efforts to understand the physical reasons behind the apparently universal density profiles found for dark matter halos in cosmological simulations (eg. Navarro, Frenk and White, 1996 (NFW); Power et al., 2003), have not quite succeeded to our knowledge. Such reasons are not to be confused with the criteria for determining the numerical validity or invalidity of a particular numerical simulation (e.g. Power et al., 2002), but are rather concerned with the collective mechanisms that may be operating (Lynden-Bell, 1967; Kandrup, 1998; Nakamura, 2000).
 
Recently two papers have addressed the relaxation problem theoretically from different perspectives. 

Henriksen (2004; henceforth I) has developed a coarse graining procedure that allows an analytic prediction of the equilibrium state of the system in the intermediate asymptotic region. Edge behaviour is also reasonably well understood as a Keplerian limit, but the  power law density profile that is reasonable in the intermediate domain is far too steep  to explain either the simulated or the observed profiles at the centres of these systems. These steep power law profiles are argued to represent isolated singular states associated with the isothermal and polytropic distribution functions (DF). Breaking the self-similar state of the halo was shown to lead to a central flattening of the density profile in time, and it was suggested that this flattening continues until a central core is established. However profiles without cores such as the NFW style cusps that are nevertheless flatter than self-similar profiles, represent from this perspective intermediate states that are less relaxed than polytropes or the  non-singular isothermal profile.  

Binney (2004) has analyzed a collection of planar `particles', each with one degree of freedom  perpendicular to the plane and has followed the development semi-analytically. That is, each orbit is defined analytically but the particles are advanced numerically. This problem is analogous to the  spherical Fillmore and Goldreich (1984) calculation. He has argued that the exact collisionless density profile is strictly singular at the centre and hence an arbitrarily large number of particles is required to resolve it. He gives both the power law density profile and the corresponding DF in the asymptotic region of the model. Extrapolating to multi-dimensional phase space and by accepting the NFW central density profile for cosmological dark matter halos, he is able to suggest a specific resolution limit for a given number of particles, that should apply generally. He attributes the entire relaxation to phase mixing in a time dependent potential. 

We agree with many of these conclusions for the toy model, which we rederive below using the coarse-graining technique. We do not agree however that the large scale collective behaviour (i.e. phase mixing) is the only relaxation mechanism in real systems.  

We identify a relaxation mechanism in addition to phase mixing that is stronger in a 3D system than in the 1D toy model. This mechanism is collective and functions above the scale of the mean inter-particle (or inter-clump) distance, being really due to short-wavelength (below the Jeans length)  wave-particle interactions known as Landau damping. 

We see these damped waves as being the mechanism (analogous to collisions in a normal thermal gas) by which energy is redistributed to the individual particles  from the large-scale collective energy sharing. This latter  process may be accurately termed `violent relaxation' (Binney, ibid). However  this type of relaxation is really towards the virialized state which is also associated with the emergence of a self-similar phase. This large-scale process might  perhaps better be termed  `violent virialization', since the Jeans length in these outer regions does not approach the interparticle scale. We expect that the damped waves with wavelength less than but close to the Jeans length, should be the most effective. 

We define an isolated system as one that undergoes no substantial dissipative mass exchange with its surroundings (either gain or loss) so that it is necessarily of finite mass. Such a system may be all of the particles that will eventually detach from the Hubble flow and virialize, or indeed  it may be any finite set such as in the 1D calculation. Any virialized system may be isolated only episodically of course.
  
We will distinguish   isolated from non-isolated systems based on the insights afforded by comparing these two papers. Once the system is virialized  and isolated  the distribution of energy to smaller scales can still be arrested (e.g. Merrall and Henriksen, I), so that  systems  cease evolving microscopically. We find that this can be due either to the rapid damping of the small scale waves or to the ordering constraint imposed by a conserved quantity in addition to the energy. In real systems the sub-scale relaxation has to be restarted by breaking the isolation in some way. Simulated systems will only be valid according to this view on scales larger than the smallest continually resolved Jeans length.

 
\renewcommand{\textfraction}{0}
\renewcommand{\topfraction}{1} 
\renewcommand{\bottomfraction}{1}

\section{Coarse-graining planar particles} \label{sec:1D}

In this section we rederive Binney's results for a 1D toy model using the methods of I. We show in addition that these results are characteristic of a self-similar regime, so that in a sense not even the NFW partially relaxed state has evolved. Since there is no angular momentum in this example and since the phase mixing long associated with the virialization is present, we explain the delayed relaxation mainly  in terms of the relative weakness of the small-scale damping. 
 
A collection of planar particles treated statistically may be described by the following principles:

{\bf 1.}
There is a mass distribution function $ F$ defined so that the normal mass distribution function $f$ is 
\be 
f(\vec v,\vec r;t)=F(v_x,x;t)\delta(v_y)\delta(v_z),\label{DF}
\ee
so that the spatial density is 
\be
\rho=\int~dv_x~F. \label{density}
\ee

The `particles' are then sheets moving parallel to the x axis although they are smoothly distributed in the statistical limit. 

{\bf 2.}
The collisionless Boltzman equation becomes
\be
\partial_t~F+v_x\partial_x~ F-\partial_x~\Phi~\partial_{v_x}~F=0,\label{CBE}
\ee
where $\Phi$ is the mean field gravitational potential.

{\bf 3.}
The potential satisfies Poisson's equation in the form
\be
\frac{d^2\Phi}{dx^2}=4\pi G\rho,\label{Poisson}
\ee
where the time dependence is understood.

Following the notation and technique of I we transform these equations according to
 \bea
 X= x e^{-\delta T},~~ Y= v_x e^{-(\delta-\alpha)T}&,~~ \Psi&=\Phi e^{-2(\delta-\alpha)T},\nonumber \\
P = F e^{(\da+1)\alpha T},~~ \Theta=\rho  e^{(2\alpha )T}&,& e^{\alpha T}=\alpha t .\label{transformations}
\eea
to obtain from (\ref{CBE}) and (\ref{Poisson}) respectively
\be
-t\partial_t~P+(1+\da)P+(\da X-Y/\alpha)\partial_X~P+((\da-1)Y+(1/\alpha)\partial_X~\Psi)\partial_Y~P=0,
\label{CBET}
\ee
and
\be
\frac{d^2\Psi}{dX^2}=4\pi G\Theta,\label{PoissT}
\ee

where 
\be
\Theta=\int~P~dY.\label{Density}
\ee

By applying a coarse-grained expansion in inverse powers of $\alpha$ we arrive, following the technique in I, at the coarse-grained solution that is exact and steady at all scales above that on which the DF is `smooth' in physical space (denoted by a subscript `o' below). This solution is found in the region where the self-similarity remains unbroken and it takes the form:

\bea
F_o=&C(E_o)^{-(\frac{1+\ad}{2(1-\ad)})},~~E_o\equiv \frac{v_x^2}{2}+\Phi_o,\nonumber \\
\Phi_o=&\gamma |x|^{2(1-\ad)},~~~~~~~~~~~~~~\rho_o=\rho_o(1)|x|^{-2\ad},\label{solC}
\eea
where the modulus sign is used for clarity although it is not strictly necessary.

The quantities $\gamma$ and $C$ are defined (after a straightforward but tedious integration of the zeroth order equation (\ref{Density}))  by 
\be
\gamma\equiv \frac{2\pi G\rho_o(1)}{(1-2\ad)(1-\ad)}
\label{gamma}
\ee
and 
\be
C\equiv \sqrt{2/\pi}\left(\frac{\Gamma(\frac{1+\ad}{2(1-\ad)})}{\Gamma(\frac{\ad}{1-\ad})}\right)\left(\frac{(1-2\ad)(1-\ad)}{4\pi G}\right)\gamma^{\frac{1}{1-\ad}}.\label{C}
\ee

In these formulae $\rho_o(1)$ is the density at $|x|=1$ and $\Gamma()$ is the usual gamma function of its argument.

The quantity $\ad$ is a free parameter but it's choice fixes a global constant whose time and space dimensions are in this ratio. We note that it should be $\ad<1/2$ in order to avoid a mass singularity at $x=0$, and in fact to maintain a force acting toward the origin.

 One obtains the results of Binney's `toy' calculation by setting \be
\ad=1/4,\label{Binney}
\ee
 since then $\rho_o\propto x^{-1/2}$ and $F=C~E_o^{-\frac{5}{6}}$ as per Binney's results. 

Moreover in this case  
\be
\gamma\equiv \frac{16\pi G\rho(1)}{3},
\ee
and 
\be
C=\sqrt{\frac{2}{\pi}}\frac{3}{32\pi G}\frac{\Gamma(5/6)}{\Gamma(1/3)}\gamma^{4/3}.
\ee

Any global constant for which $\ad$ is between zero and $1/2$ is however acceptable in the coarse-graining, and these possibilities correspond to different imposed global constants on the self-similarity. If for example the initial condition were not one of an isolated system, but rather a density distribution according to a  power law of the form $\rho\propto x^{-\epsilon}$, then one imposes (at least over a limited range) the global constant of dimensions $ML^{\epsilon-3}$, from which on eliminating $M$ using the dimensions of $G$ we obtain a constant of dimensions $T^2/L^{\epsilon}$. This corresponds to $\ad =\epsilon/2$. We would thus obtain the same behaviour as above if $\epsilon =1/2$, which corresponds to starting the system in the self-similar state and allowing it to stay there indefinitely as matter continues to fall in. This is not an isolated system. 

However the initial condition used by Binney is not a power law but is in fact 
homogeneous. So the reason for the spontaneous development of the self-similar behaviour may appear mysterious. But any quantity that is asymptotically determinant, and which has dimension $T^4/L$ (or an appropriate power thereof, so that $\ad =1/4$) is a candidate for dictating the emergent self-similarity. 

Binney's initial conditions contain $G\Sigma$ ($\Sigma$ is the surface density  so that the product has dimensions $L/T^2$) and since the system is isolated (there is a finite number of particles) we expect the phase space volume $J$ of dimension $L^4/T$  to be conserved. From these quantities we can construct a characteristic scale and a characteristic time and a characteristic velocity from their ratio as $V=((G\Sigma)^3J)^{1/7}$. The existence of these conserved scales is consistent with assigning an initial time and location as is done in the toy model. These prevent any initial self-similarity but should become less important asymptotically. The phase space volume and the velocity are not merely initial values however and may be expected to dictate the eventual self-similarity. Together the two form a  quantity $J/V^5$ of required dimension $T^4/L$. This apparently arbitrary combination is unique in that it interchanges $L$ and $T$ in the phase space volume. We can not on dimensional grounds alone distinguish between $\ad=1$ dictated by the constant velocity $V$ and $\ad=1/4$ dictated by the constant $J/V^5$. However the requirement that $\ad<1/2$ excludes the first possibility.  A non-isolated system  such as is the case for our self-similar model, does not necessarily follow this development, since  the phase space volume is not conserved and the constant associated with the initial density profile becomes determinant.

 However we have seen that the self-similar regime in Binney's model is in fact  equivalent to the continuing accretion of an $x^{-1/2}$ power law. Nevertheless despite the prediction (I) that the self-similar regime should be subject to central flattening, there is little evidence of flattening in the 1D calculations (figure 2, Binney, 2004). 
 Thus there is a  need for a reconciliation that may afford some insight into how the basic mechanisms depend on accessible phase space volume. 
This is discussed in the next subsection with appeal to a technical appendix.

\subsection{Dependence of relaxation on phase space dimensionality}

In appendix A we observe first that the zeroth order coarse-grained self-similar profile in 1D remains valid to the centre of the system, since the gravitational acceleration does not diverge there. Thus we can  expect  a less evident central flattening than was found in 3D as an immediate result of the change in dimensionality (cf Teyssier et al., 1997).

 However if the flattening  is ultimately due to small scale collective wave-like relaxation , then we  expect to detect a difference between the three dimensional system and the present example by fine-graining. This has been carried out in appendix A where the first order fine-grained gravitational acceleration is shown to satisfy 
\be
 X^2\frac{d^2g_{-1}}{dX^2}-(1-\ad)X\frac{dg_{-1}}{dX}+2\ad(1-2\ad)g_{-1}=0.\label{FGA} 
\ee
analogous to equation (48) in I. This is an Euler equation with solution
\be
g_{-1}=A_{-1}X^{p_+}+B_{-1}X^{p_-},\label{solg-1}
\ee
where 
\be
p_{\pm}=1-\frac{1}{2}\ad\pm\sqrt{(1-\frac{1}{2}\ad)^2-2\ad(1-2\ad)}.
\label{indices}
\ee
For $\ad=1/4$ this gives $p_{\pm}=(7\pm\sqrt{33})/8$ or about $1.593$, $0.157$ respectively. Unlike the Bessel function behaviour revealed in the equivalent study of the three dimensional case, {\it the variation with $X$ is asymptotically monotonic, possessing at most a simple extremum if the constants are of opposite sign}. This is moreover true for all values $0\le \ad\le 1/2$. Since this simple structure is a function of $X=r/(\alpha t)^{\da}$ it may be regarded as a soliton wave, but it is on a scale comparable to that of the system.  

Higher order terms such as $g_{-2}$ are produced by satisfying the same homogeneous equation (\ref{FGA})  forced by a linear combination of powers. This will  introduce more wave-like sub-structure as a function of $X$, but it is progressively weaker and remains more soliton-like than wave-like. It remains a sum over positive powers so that no divergence occurs at the origin. 

In contrast the higher order fine-grained terms in the three dimensional problem are produced from the homogeneous equation forced by Bessel functions. This creates sub-structure that is formed from the products of Bessel functions of various phases so that the higher order fine-graining  detects a `cascade' to waves of ever shorter wavelength and weaker amplitude. This cascade is much less pronounced in the  one dimensional problem.     
   
We commonly refer to this sub-structure as a superposition of `propagating waves' since it is a function only of $X(r,t)$, but this structure includes the phase sheet winding due to phase mixing as well as collective effects on a smaller scale. During the self-similar accretion the phase sheet spiral does propagate outwards (e.g. HW99). At a fixed $X$, that is moving with the wave, the associated density $\rho_{-1}$ is damped as $1/t^2$. At a fixed $x$ there is generally a damped mode ($p_+$) and a growing mode ($p_-$), but if the $p_-$ mode is excluded by requiring finite behaviour at $x=0$ then only the damped mode remains.This is similar to what has been found for the spherical isotropic sub-structure as presented in paper I (equation (53)), but as has been already remarked the waves are more prominent in that case.  

We conclude that the rapidity of central flattening  that appears in the final coarse-grained system when the self-similarity is broken, is correlated with the prominence of the  wave-like behaviour of the fine-grained structure. This time dependence includes the phase mixing in a changing gravitational field as in Binney (ibid), which has been shown to be sufficient to establish the coarse-grained self-similarity. However  there is in the fine-graining analysis an evident cascade of structure to smaller and smaller scales with associated damping.  

Thus collisionless relaxation takes place in a progressive manner from large to small scales. Virialization represents relaxation on the largest scale and it is `violent' in the sense that it depends on the phase mixing in the time-dependent gravitational potential associated with large amplitude fluctuations. This `violent virialization' happens rapidly at the beginning of the self-similar accretion regime (e.g.  Natarajan, Hjorth and van Kampen, 1997;HW99; LeDelliou and Henriksen, 2003; Binney (ibid)) and is responsible for its form. However the central flattening of the density profile halo inferred in I requires a self-similarity breaking mechanism, which we attribute  here to a longitudinal wave cascade.

Such waves are subject to Landau damping below the Jeans length as they transfer collective energy to the dark matter `particles'. Dynamical friction may be attributed to the emission and subsequent damping of these waves (Marochnik,1968) This is the mechanism that we nominate as the analogue of collisions in a normal thermal gas, which we expect to lead ultimately to a maximum entropy configuration.

The relevant Jeans length here is $\lambda_J\equiv \sqrt{\frac{\pi \sigma^2}{G\rho_o(x)}}$, where $\sigma^2$ is the local velocity dispersion which in the self-similar regime, varies with $\Phi_o$. So in the 1D case $\lambda_J\propto x$ and vanishes as $x\rightarrow 0$. This is also true for the higher dimensional cases discussed in I, where $\lambda_J\propto r$. Thus central similarity-breaking relaxation must be due to rather short-wavelength Landau damping.  

In reality the cascade must stop at the interparticle scale $n^{-1/3}$, where $n$ is the particle density if the dark matter particles are distributed uniformly, but would be the clump density if the particles are clumped. An example of such clumping would be the `mini-halos' of Diemand, Moore, and Stadel (2005), who suggest that near the sun $n\approx 500/pc^3$, so that the minimum scale would be $\approx 1/8$ pc. In either case the mean field distribution function approach requires minimum volumes that contain many particles. Thus (assuming Gaussian fluctuations) if $s$ is the fluctuation percentage of the total number of particles ($1/s^2=nL^3$) in a volume $L^3$, then effectively the mean-field cut-off scale is $L=1/(s^2n)^{1/3}$. For $s$ as large as $1\% $, this can be some $20$ times the interparticle scale. Below this scale even the definition of a mean density is problematical. 

The ratio of the Jeans length to the interparticle scale goes to zero more rapidly in the 1D system ($\propto x^{1/2}$) than it does in the spherical system ($\propto r^{(1-\frac{2}{3}\ad)}$, $\ad>1$). It would therefore be more difficult to resolve the 1D relaxation scale than the 3D scale in a mean field (averaged) study.  

We remark that although the amplitude  and the energy of these waves decreases with scale,they actually transfer their energy more rapidly since the Landau damping time varies as $\approx \lambda_J/\sigma\approx 1/\sqrt{G\rho_o}$ for short wavelengths. In the self-similar regime this also decreases with scale.  It is therefore likely that these waves must be continually restarted to effect the relaxation. {\it This is a significant difference with collisional relaxation in an isothermal gas, and renders  collisionless relaxation much more contingent}.

Because the Jeans length decreases near the centre of the system in the self-similar regime that is established early in the evolution, the Landau damping relaxation will become harder and harder to resolve as the centre is approached.  The simulations of necessity always involve a finite number of particles and a finite time. There is thus a limit to the relaxation they can resolve both in time and in space. A cut in resolution that is above the local Jeans length will effectively hold the simulated system in an intermediate (between self-similar and wholly relaxed) state. However these waves may also disappear in reality precisely because of their damping efficiency, if they are not continually fed energy from larger scales.  It is inevitable that this occurs first near the centre of the system where the damping time is shortest.
 
Consequently it is not surprising that N-body systems may be trapped in intermediate states with subsequent relaxation contingent on various means of breaking the system isolation. Why this intermediate state should so frequently be the NFW cusped state in simulations and perhaps in reality (e.g. Pratt and Arnaud, 2003) will be discussed further below.
      
\subsection{Independent Support for small scale relaxation}

There is support of various kinds for the suggestion of Landau damping relaxation in the literature.  Kandrup and Novotny (2004) have suggested independently on numerical grounds that systems may depend on wave-like perturbations in order to reach a true equilibrium. Without these perturbations they find also that the system may become `frozen' in a kind of false equilibrium.

Such wave-particle `relaxation' was also  suggested earlier on numerical grounds by Funato, Makino and Ebisuzaki (1992a,b) and was inferred in Merrall and Henriksen (2003) where they studied virial oscillations in some detail. In the latter paper it was demonstrated explicitly that the approach to virial equilibrium was much smoother as the number of particles in the simulation increased so that shorter and shorter wavelengths were resolved. This was in fact interpreted as reflecting the presence of ever smaller spatial scales in the relaxation. 

 The vigorous oscillations that appeared with fewer particles in Merrall and Henriksen (ibid) were also physical since their amplitude exceeded $1/\sqrt{N}$. These should represent an intermediate part of the heirarchy that we propose. Such intermediate scale oscillations were also detected by David and Theuns (1987), and perceptive comments about relaxation and Landau damping have been made by Kandrup (1998).
 
A very similar argument has been used by El-Zant et al. (2004) to explain the absence of a density cusp in the cluster MS 2137-23. They appeal to dynamical friction between the dark matter and the inward spiraling galaxies to overcome adiabatic compression and flatten the cusp. But dynamical friction is in fact associated with Landau damping (Marochnik, 1968). Our proposal is that  dark matter fragments accreting on elliptical orbits around a larger halo will also inspiral as they emit longitudinal waves. These waves can restart the relaxation process at the local Jeans scale, which may be quite large in the outer regions of a halo. In fact these are probably the intermediate scale oscillations that establish the self-similar intermediate profile. It is readily verified that in the self-similar region $\lambda_J\propto X$, so that the relaxation scale and the halo both grow as $t^{\da}$.   

We note that the smallest, youngest, halos found by Diemand et al. (ibid) show cusped density profiles very nearly characteristic of the self-similar regime. That is, even though $\alpha\beta\gamma$ profiles may be fitted to the data, simple power laws between $1.5$ and $2.0$ also fit. In the lowest mass profiles (presumably the youngest) the power is close to $2$ so that  even the relaxation implied by the new NFW profile of Power et al. (ibid) is not seen. This supports the view that the self-similar virialization is the first stage in the relaxation.  

We infer that this is followed by a central evolution towards a softening cusp, under the influence of ever smaller scale collective processes. For these to go to completion at the centre and so form a flat core consistent with the coarse-grained DFs as deduced in I, the central Jeans scale must be resolved in simulations and the waves must be continually `pumped' in real systems.
Even if the Jeans length is resolved and/or  the simulation is smoothed on the smallest continually resolvable Jeans length,
the central regions of a simulation should not be isolated  if a core is to be detected.

\subsection{Why is the cusped density profile so prevalent?}
   
The preceding sections argue that the NFW cusped state is a meta-stable state of (largely central) relaxation between the virialized self-similar state and a completely flattened core. In this section we isolate what we believe to be the principal reason for this meta-stability. 

In the first instance we expect only the  oldest systems (low mass at low redshift) to show polytropic or isothermal cores because the relaxation is contingent on accretion/mergers and is faster at higher densities.  So it is not surprising that young large systems such as clusters should be in an intermediate state, except perhaps the cD galaxy. However this does not explain the existence of the NFW state to very small scales in simulated halos that have evolved over the Hubble time.   

We have  seen that the resolution of the Jeans length is essential for relaxation in simulated systems, and non-isolation of the system is essential for relaxation in both real and simulated systems. Resolution effects can not explain of course the outer power law region, but this is probably due to the very large Jeans length compared to the interparticle scale in this region. It is only in the centre where it can approach the interparticle scale. To see a relaxed outer halo one would have to smooth on  a macroscopic scale (e.g. Merrall and Henriksen (ibid).
 
One clue is provided by the argument based on isolation. Is the central part of the NFW state isolated?
 By this we mean  that the eventual central particles may be identified early in the process of halo formation. If this is the case then they  form a finite self-similar system just as in the 1D calculation by Binney (ibid). We apply therefore the same argument for the asymptotic self-similarity that we used in the first section for the 1D  and compare our results to a direct  inversion (Widrow, 2000) of the NFW density profile. 

 As before, we can expect the mass and the phase space volume to be constant. This gives $GM$ and the phase space volume  $J$ as constants with respective dimensions $L^3/T^2$ and $L^2/T$ (or the cube of this quantity). Neither one of these alone dictates the observed asymptotic central evolution, but together they form $V=GM/J$ of dimension $L/T$. From these constants there also follows a characteristic scale  and a characteristic time. However these can not give an asymptotic self-similarity and so just as for the toy  model we suggest that a combination of $J$ and $V$ in the form $J/V^3$ , which interchanges $T$ and $L$ in $J$, is the determining constant. Once again one has to choose between $\ad=1$ that follows from the constant velocity alone and $\ad=1/2$ following from the latter combination. Since $\ad=1$ is the limit towards which flat systems tend during continuing infall (e.g. HW99), we can assume that an isolated system remains in the flat state ($\ad<1$).

   In any case using $\ad=1/2$ we obtain from the general coarse-grained result of I (equation 31) that 
\be
f=C|E|^{-5/2},\label{DFNFW}
\ee
and of course that (I-35)
\be
\rho_o=\rho(1) r^{-1},~~~~~~~~~~\Phi_o=\gamma  r,
\ee
where $\gamma $ is as above except that $(3-2\ad)$ replaces $(1-2\ad)$.

In the appendix of Widrow (2000) the fitting function for NFW cusps is given as $f\propto (1-{\cal E})^{-5/2}$ as ${\cal E}\rightarrow 1$, which with an appropriate choice of the potential at infinity is $f\propto E^{-5/2}$. This holds when $E\ge 0$ which includes the most tightly bound particles.
 It is  significant that this is not a linearly stable DF by the usual theorems (e.g. Binney and Tremaine, pp 306-307). Thus temporal evolution can be  stimulated even by very weak disturbances if they are present.

 It therefore seems plausible that the NFW type cusps are characteristic of isolated systems in which small scale relaxation has damped away.  
But for this to be the case most of the halo particles must be on elliptical orbits and one must enquire into the origin of this angular momentum. 

Merrall and Henriksen (2003) found numerically that colliding halos that formed initially a wide binary did not evolve central gaussian DFs. The likely explanation is  that the centre became isolated due to the angular momentum of most of the system particles, but this is a rather special example of major mergers.

In LeDelliou and Henriksen (2003) it was shown explicitly that the presence of angular momentum restricted the phase space volume of the system dramatically. Moreover,a `Keplerian' distribution of angular momentum, similar to that found in cosmological simulations, was found to produce the NFW profile in isolated self-similar infall. But the angular momentum was introduced arbitrarily. 

These results are  consistent with the idea  that the more symmetric the system (and hence the more integrals) the less likely is the relaxation to a unique state (Aly,1989; I, appendix A). Although the `thermalized' or maximum entropy state  always remains possible, it is no longer the unique coarse-grained result. Coarse-graining theory as presented in I does not detail the inhibiting mechanism, but we note that the differential motion can supply free energy to what would otherwise be  chaotic, relaxed, motion.

The only exception to this inhibition of relaxation by angular momentum is solid body rotation. For then, in the notation of I, $\alpha=0$ because of the constant angular velocity, say $\Omega$. The coarse graining can nevertheless be carried out as in I by expanding in powers of $\delta^{-1}$. The precise argument may be deduced from the general equations (17)-(19) of I by taking $s=\ln{R}$ and multiplying expressions that contain $\da$ by $\ad$. Then by setting $\alpha=0$ (no time scaling because of fixed $\Omega$) we predict a homogeneous system (to logarithmic accuracy with a finite cut-off in velocity) with DF 
\be
f=CE^{-3/2}.
\ee
The divergence anticipated on page 223 of Binney and Tremaine(ibid) for spherically symmetric isotropic systems is avoided because the real potential is everywhere positive (in fact it is harmonic) and we truncate the velocity at a large but finite value, consistent with the spatial coarse graining ($R$ should not be taken zero). We therefore predict that relaxed systems should be in solid body rotation, including zero. It does not follow however that systems in solid-body rotation are relaxed, since the various criteria discussed above must also be met.

 Now in fact the computed cosmological halos have  what appears to be a characteristic specific angular momentum distribution (Bullock et al., 2001; LeDelliou and Henriksen, 2003) not greatly different from the `Keplerian' $\propto \sqrt{GM(r)r}$. It is important to note that the distribution does not correspond to rigid rotation unless $\rho(r)$ is constant. In the latter case $j\propto r^2\propto M^{2/3}$ rather than the $1.3\pm 0.3$ as determined by Bullock et al. (ibid). The latter is closer to the self-similar infall value of $j\propto M^\mu$ where $\mu=(2-\ad)/(3-2\ad)$ (Le Delliou and Henriksen, 2003) as might be expected.

The origin of this angular momentum  is generally thought to be  either tidal in the linear regime or the result of mergers or possibly both (Maller, Dekel and Somerville, 2002). However recent simulations by MacMillan  (MHW05) find the same angular momentum distribution within errors {\it even when the entire simulation is limited to a single large halo and the BBKS initial conditions  initial conditions are removed}. This seems to be an effect of the radial orbit instability, although the amplitude of the bulk angular momentum is very small. Restoring the BBKS initial conditions produces a larger mean square value, although it is still an order of magnitude below the full cosmological mean spin parameter. This value presumably requires the presence of neighbouring halos. Nevertheless the distribution of angular momentum remains the same to within errors.  Thus it seems  inevitable that the angular momentum distribution associated with the NFW meta-stable state should arise during the formation of even an isolated halo, although the amplitude depends on sub-structure and the environment.    
For the isolated system the bulk angular momentum is divided equally between particles inside and outside the virial radius.
Gradually these satellite particles should join the virialized system, although they may have sufficent angular momentum so as not to approach the isolated centre. Those that do pass directly through the centre will have essentially zero angular momentum and will not change the angular momentum of the central state. It is the satellites that spiral in gradually under the influence of tidal friction that should eventually relax the central regions by destroying the isolation. We remark that if this effect is not correctly contained in the simulations then both an isolated centre and an overabundance of satellite systems would result. In the outer regions we can expect counter-rotation to persist in real and simulated systems, although it is possible that relaxation forms two counter-rotating solid body systems. A cluster of galaxies is a young relatively isolated system so that one expects an NFW profile (Pratt and Arnaud, 2003) together with this counter-rotation between virialized and unvirialized galaxies. In particular the central cD may be counter-rotating relative to the average of more distant members of the cluster. 

This single system evolution is to be contrasted with the path to equilibrium followed by a non-isolated real system that is evolving by merging and the accretion of substantial fragments.  In such a case the net angular momentum of the system may be decreased or increased depending on the collision prameters. In either event the local Jeans length would be increased with the mean square velocity dispersion $\sigma^2$, and the cascade to the Landau damping scale would also be restarted, appearing as dynamical friction. When the net angular momentum is reduced  one could expect that the evolution towards a central, solid body, thermalized state would continue. Simulated mergers must resolve the central Jeans length to see this effect. The intermediate scale waves are responsible for self-similarity and virialization, and are usually well resolved.

So the answer to the title question of this section is: central cusps result from  the rapid dissipation of sub Jeans length relaxation combined with the isolating effect of differential rotation with which cusped profiles are always associated. The failure of the simulations to remove these cusps entirely during the secondary infall of satellites and merging  is probably due to the failure to resolve and/or maintain the Landau damped waves. Moreover nearly circular orbits decaying slowly by dynamical friction are the likely best catalyst for complete relaxation. The NFW cusps are already relaxed relative to the self-similar cusps.

\section{Discussion and Conclusions}

The basic conclusion of this paper is that small scale Landau damping is responsible for the ultimate `thermalization' of a collisionless dynamical system. Moreover this collective behaviour damps more rapidly at small scales and has to be restarted in order to drive the system to `thermal equilibrium' (maximum entropy state). Both isolation and differential rotation inhibit the necessary `restarting'. 

 Isolated systems, both real and simulated, can be frozen into an intermediate state between self-similar infall that is established by violent virialization, and a flattened thermalized core. The NFW profiles, old and new(Power et al., 2003), are of this form. 

It is  vitally important that a simulation retain sub-Jeans scale resolution  if it is to be followed into a central relaxed state. This is equivalent to correctly incorporating dynamical friction there. Conversely, if a given simulation is smoothed on a Jeans scale that is fully resolved throughout the simulation, then we might expect to see a profile with a flattened core. Such smoothing was effectively carried out in Merrall and Henriksen (2003). 
   
As an additional test of our ideas a referee has called our attention to the pertinent analysis of the Draco and Ursa minor dwarf spheroidals (Kleyna et al., 2002; Kleyna et al., 2003). 

In the case of the Ursa minor {\bf dsph} there is indirect evidence (Kleyna et al., 2003) for a core of radius $r\approx 500pc$. The question arises as to what constraints this fact, together with our considerations, might place on the entire dark matter halo. According to the isothermal flattening model of I, we can expect flattening on a scale $r$ after a time $t$ provided
\be
\sqrt{GI_{oo}}\frac{t}{4r}\approx 3/4.
\ee
For an isothermal halo $I_{oo}=M_h/(4\pi r_h)$ and so with $r=500pc$ and $t$ set equal to the collapse time at the epoch $z$, namely $0.543/\sqrt{G\rho_b(z)}$, we obtain a constraint on the ration of halo mass to radius $M_h/r_h$. With 
$\rho_b=0.27\times 1.88\times 10^{-29}(1+z)^3$ and $z=9$, this constraint becomes
\be
\frac{M_h}{r_h}\approx 7\times 10^6 M_\odot/kpc.
\ee
This is too small a concentration since for example if the halo mass is as much as $10^9$ $M_\odot$, the halo radius must be as large as $\approx 140kpc$ !
However, if the effective relaxation time for the core were shorter than the background collapse time by only a factor $3$ or so (due to detachment from the Hubble flow), this scale would be decreased by an order of magnitude. The same result obtains if the size of the core is reduced by a factor $3$. 

Although the Draco {\bf dsph} plausibly has an isothermal, nearly isotropic, halo (Kleyna et al., 2002), it shows no evidence of a core down to $\approx 100pc$. This implies a more reasonable concentration for Draco and suggests that the size of the Ursa minor core may be overestimated. 

 If near the centre of Draco the dispersion is indeed $\sigma\approx 8.5$ km/sec, and if  the density is about $10^{-22}$ $ g/cm^3$ (which is the value at $100$ pc if the galaxy has a mass of $10^9M_\odot$ in an isothermal distribution at an epoch $z=9$), then the central Jeans length is $\approx 190pc$. This can be taken as a reasonable upper limit to the size of a core given the apparent isothermal outer behaviour and our arguments concerning relaxation. Such cores should have nearly solid body rotation (including zero).

\section{Acknowledgements}

This work was supported in part by a Discovery grant from the canadian NSERC. The basic ideas were developed thanks to the hospitality of the Service d'Astrophysique, CEA Saclay. The author wishes to thank the referee for very pertinent criticism that led to substantial improvement in the manuscript.

\newpage

\appendix{{\bf Appendix A:  Flattening and Fine-graining the one dimensional model}}
\vskip 1 true cm
\renewcommand{\theequation}{A\arabic{equation}}
We know that the coarse-grained self-similarity can not continue to the centre of the system in the three spatial dimensions problem (Henriksen, 2004) since the gravitational acceleration diverges there (for $\ad>1$) and the coarse-graining expansion must fail. A glance at equations (\ref{solC}) shows that this is not the case ($\ad<1/2$) for the one-dimensional problem. Consequently we can expect the coarse-grained self-similarity to continue to the centre with the attendant density power law, in accordance with the numerical conclusions of Binney (ibid). 

However there remains the problem of how the system will react to temporal perturbations. This may be discussed  by restoring the time dependence to equation (\ref{CBET}) which breaks the self-similarity. A long calculation of the kind given in Henriksen(ibid) shows that in fact the first order density term grows {\it relative} to the zeroth order term as $X^{\ad-1}$, and that it is negative in sign so that it tends to flatten the cusp. Compared to the three dimensional results this is however rather slow, since in that case one finds the relative flattening to go as $R^{-1}$ in the isothermal limit and as $R^{-3+\ad}$ in the case of infinite polytropes ($\ad>5/4$). We recall that for the present example $0<\ad<1/2$, $X=x/(\alpha t)^{\da}$, and that the system is non-isolated. In spherical geometry $R=r/(\alpha t)^{\da}$.  

Thus we expect flattening eventually given the presence of perturbations and sufficient time at a finite $x$, but it is generally slower to appear. The limitation on the perturbation growth rate is compounded by the persistence of the zeroth order self-similar behaviour to the centre of the system.

We turn now to a fine-graining of the system to see if the one dimensional example is unique at this level. Once again the calculation is similar to that described in Henriksen (2004), but we will give it here in detail since it is intimately connected to our argument in the text.

We proceed by accepting the  coarse-grained solution in the more convenient form 
\bea
 P_o=P_{oo}(\zeta)X^{-(1+\ad)},&~~~~~& P_{oo}=C\epsilon_{oo}^{-(\frac{1+\ad}{2(1-\ad)})},\nonumber\\
\zeta\equiv YX^{-(1-\ad)},&~~~~~& \epsilon_{oo}\equiv \frac{\zeta^2}{2}+\gamma,\\
\Psi_o=\gamma X^{2(1-\ad)},&~~~~~& \Theta_o=X^{-2\ad}\int~P_{oo}~d\zeta,\nonumber
\eea
where the integral in $\Theta_o$ is generally denoted $I_{oo}$ and is numerically equal to $\rho_o(1)$. $C$ is as given in the text.

We fine grain about this solution by using the expansion 
\be
P=P_o+P_{-1}\alpha +P_{-2}\alpha^2+\dots,
\ee
as $\alpha\rightarrow 0$, in equation (\ref{CBET}) without the time dependence.
In this way we expect to detect the phase mixing that is part of the process of the self-similar relaxation. Another part may be due to collective wave-like processes.

From the zeroth order we obtain on remembering the equation satisfied by the zeroth order DF
\be
Y\partial_XP_{-1}-\partial_X\Psi_o\partial_YP_{-1}-\partial_X\Psi_{-1}\partial_YP_o=0,\label{firstorder}
\ee
which we proceed to solve by characteristics. The only characteristic constant is 
\be
\epsilon=\epsilon_{oo}X^{2(1-\ad)},
\ee
and along the characteristic we find
\be
\frac{dP_{-1}}{ds}=\frac{d\Psi_{-1}}{dX}\partial_YP_o.
\ee
After computing the partial derivative on the RHS of this equation and noting that $dX/ds=Y$, we obtain
\be
\frac{dP_{-1}}{dX}=\frac{d\Psi_{-1}}{dX}X^{-(3+\ad)}\frac{P_{oo}}{\epsilon_{oo}}\frac{1+\ad}{2(\ad-1)}.\label{charP-1}
\ee
 An integration of this equation along a characteristic would produce in general an arbitrary additive function $F_{-1}(\epsilon)$. However at $t=0$ or $x\rightarrow\infty$ so that $X$ is large we will require $P_{-1}$ to vanish (no relaxation) on all characteristics so that we set this function equal to zero.

We  use  the first order Poisson equation (\ref{PoissT}) together with (\ref{Density}) to obtain 
\be
4\pi G\int~P_{-1}~dY=\frac{d^2\Psi_{-1}}{dX^2},
\ee
and then after differentiation with respect to $X$ and using  the result (\ref{charP-1}) as true everywhere
we find after some algebraic rearrangement that

\be
X^2\frac{d^2g_{-1}}{dX^2}-(1-\ad)X\frac{dg_{-1}}{dX}+2\pi G I\frac{(1+\ad)}{(1-\ad)}g_{-1}=0.\label{g-1}
\ee   
where $g_1\equiv -d\Psi_{-1}/dX$ and 
\be
I\equiv \int~\frac{P_{oo}}{\epsilon_{oo}}~d\zeta.\label{Iint}
\ee
This integral can be evaluated explicitly by combining $\rho_o(1)=I_{oo}$ in equation (\ref{gamma}) with the integral evaluation of $I_{oo}$  and the value of $C$ from equation (\ref{C}) to obtain the simple result

\be
2\pi G I =2\ad(1-2\ad)(\frac{1-\ad}{1+\ad}).\label{I}
\ee
In the course of the calculation we make use of 
\be
\int_0^\infty~\frac{du}{cosh(u)^{\mu/2}}=\frac{\sqrt(\pi)}{2}\frac{\Gamma(\mu/2)}{\Gamma((1+\mu)/2)},
\ee
provided that $\mu>0$.

This gives finally equation (\ref{g-1}) as 
\be
X^2\frac{d^2g_{-1}}{dX^2}-(1-\ad)X\frac{dg_{-1}}{dX}+2\ad(1-2\ad)g_{-1}=0,
\label{fg-1}
\ee 
which is discussed in the text.

\label{lastpage}
  
 \end{document}